\documentclass{article}
\usepackage[english]{babel}
\usepackage{graphicx}
\usepackage{amsmath, amssymb,xcolor,graphics,epsfig,amsbsy,latexsym, amsthm, amsfonts} 
\usepackage{hyperref}
\usepackage{apacite}
\usepackage{natbib}
\usepackage{tabularx,subfig,enumitem}
\usepackage{ragged2e}
\usepackage{color,soul, setspace}
\usepackage[letterpaper,top=2.54cm,bottom=2.54cm,left=2.54cm,right=2.54cm,marginparwidth=2.54cm]{geometry}
\usepackage[font=small, skip=0mm]{caption}
\usepackage{multirow}
\usepackage{footnote}
\usepackage{url}
\usepackage{booktabs}
\usepackage{adjustbox}
\title{ A Joint Model of Longitudinal CVD Risk
Factors, Medication Use, and Time-to-
Terminal Events}
\author{Zeynab Aghabazaz$^{1*}$,
Michael Joseph Daniels$^{2}$,
Donald M. Lloyd-Jones$^{3}$,
Juned Siddique$^{1}$ \\
$^{1}$ \footnotesize Department of Preventive Medicine, Northwestern University Feinberg School of Medicine, Chicago, IL, USA \\
$^{2}$ \footnotesize Department of Statistics, University of Florida, Gainesville, FL, USA \\
$^{3}$ \footnotesize \parbox{15cm}{Framingham Center for Population \& Prevention Research and Section of Preventive Medicine \& Epidemiology, 
Department of Medicine, Boston University Chobanian \& Avedisian School of Medicine, Boston, MA, USA} \\
\footnotesize *\texttt{zeynab.aghabazaz@northwestern.edu}}
\date{}
\begin{document}
\maketitle
\begin{abstract}
We introduce a novel Bayesian approach for jointly modeling longitudinal cardiovascular disease (CVD) risk factor trajectories, medication use, and time-to-events.
Our methodology incorporates longitudinal risk factor trajectories into the time-to-event model, considers the temporal aspect of medication use, incorporates uncertainty due to missing medication status and medication switching, and analyzes the impact of medications on CVD events. Our research aims to provide a comprehensive understanding of the effect of CVD progression and medication use on time to death, enhancing predictive accuracy and informing personalized intervention strategies.
Using data from a cardiovascular cohort study, we demonstrate the model's ability to capture detailed temporal dynamics and enhance predictive accuracy for CVD events.
\end{abstract}
\noindent%
{\it Keywords:}
Bayesian methods; Longitudinal data; Missing data.
\section{Introduction}
Cardiovascular disease (CVD) is the leading cause of death in the United States and worldwide, responsible for over one-third of all annual deaths. The onset and progression of CVD occurs gradually throughout life, beginning in early adulthood and increasing in prevalence and severity with age. Understanding how CVD risk evolves over time is essential for improving prevention and treatment strategies.
Importantly, cumulative exposures and long-term trajectories of these factors over decades are significantly linked to subsequent CVD risk \citep{allen2014blood, pool2018use, domanski2023association}. These findings underscore the necessity of longitudinal dara when evaluating CVD risk and developing intervention strategies.

Medication plays a pivotal role in modifying CVD risk factors, thereby influencing the likelihood of future clinical CVD events. Incorporating medication status into the modeling of mean trajectories allows for a better understanding of how medication initiation influences the trajectories of risk factors over time. Such insights are valuable for evaluating the timing and duration of treatment interventions and their effects on the biological processes underlying risk factor dynamics. Capturing these temporal relationships enables the development of more precise and nuanced treatment strategies, optimizing outcomes for individuals at risk of CVD.

For example, cholesterol-lowering medications, such as statins, have been shown to significantly reduce CVD risk by lowering low-density lipoprotein (LDL) cholesterol levels. Evidence from the Cholesterol Treatment Trialists’ (CTT) collaboration demonstrates that a 1 mmol/L reduction in LDL cholesterol through these therapies reduces major vascular events by approximately 20\%, irrespective of baseline LDL levels \citep{trialists2010efficacy}. These benefits extend across diverse populations, including those at lower baseline risk, underscoring their potential for both primary and secondary CVD prevention \citep{mihaylova2012effects}.

While effective at the population level, the temporal dynamics of cholesterol-lowering medication use, such as initiation, adherence, and discontinuation, are less understood at the individual level. Medication effects are inherently time-dependent, with immediate benefits, like rapid LDL cholesterol reduction upon initiation, and adverse outcomes, such as increased risk following discontinuation. Understanding these time-dependent effects is essential for elucidating the broader role of these therapies in CVD progression.

We introduce a novel Bayesian methodology for jointly modeling longitudinal trajectories of CVD risk factors, medication use, and time-to-event outcomes, utilizing data from a longitudinal cohort study. Specifically, this study
\begin{enumerate}[leftmargin=5mm]
\item[i)] Establishes explicit connections between risk factor trajectories, medication initiation or discontinuation, and clinical events, highlighting their collective influence on CVD progression.
\item[ii)] Employs methods to effectively handle missing medication status, and addresses uncertainties associated with changes in medication use between study visits.
\item[iii)] Leverages features derived from an individual's longitudinal risk factor trajectory to enhance the predictive power of medication and event-time models, enabling more accurate and personalized intervention strategies.
\end{enumerate}

When dealing with missing time-varying medication data in time-to-event models, a last observation carried forward (LOCF) method is commonly used, which assumes that the last observed medication status remains the same until the next observation, even if there is a gap or missing data \citep{shao2003last, fitzmaurice2012applied}. In this manuscript, we adopt a more refined approach, arguing that the treatment of missing medication data, and the gaps between known medication use, should depend stochastically on the specific ages and risk factor features present during the interval between collected visits, rather than relying deterministically on the last previously observed status. 

The remainder of this manuscript is structured as follows. Section~\ref{sec2} provides an overview of the Multi-Ethnic Study of Atherosclerosis (MESA), a longitudinal CVD cohort study of diverse racial and ethnic groups that motivates this work. Section~\ref{sec3} introduces our joint model framework, focusing on the integration of risk factors, medication use, and event dynamics, as well as approaches for handling missing data and uncertainty. 
Section~\ref{sec4} presents the key results from the analysis of the MESA data, and Section~\ref{sec5} concludes with an exploration of the findings' significance and potential future applications.

\section{Description of the MESA Cohort and Key Variables} \label{sec2}
Our data come from the MESA cohort, a comprehensive cardiovascular cohort study sponsored by the National Heart, Lung, and Blood Institute (NHLBI) of the National Institutes of Health (NIH)~\citep{bild2002multi}. MESA, which began in 2000, officially enrolled 6,814 men and women aged 45 to 84 years from six communities across the United States, with participants recruited from diverse racial and ethnic backgrounds. Participants have been followed for over two decades, with examination visits typically occurring every 2-4 years. The six study sites include Columbia University (New York), Johns Hopkins University (Baltimore), Northwestern University (Chicago), UCLA (Los Angeles), the University of Minnesota (Twin Cities), and Wake Forest University (Winston-Salem).

MESA is notable for its diversity and large-scale longitudinal design. To date, participants have attended up to five study visits, during which detailed clinical, demographic, and lifestyle data were collected. For this analysis, we focus on a subset of MESA participants aged 65–69 years at baseline. 
Table \ref{Demo} provides a summary of key demographic characteristics of this subset of the MESA cohort, highlighting differences by sex and race/ethnicity. The study sample included approximately 39\% White, 28\% Black, and 33\% individuals classified as Other racial/ethnic groups among men, with similar proportions among women. Educational attainment also varied, with a majority of participants reporting education beyond high school.
\begin{table}[!t]
    \centering
    \caption{Demographic characteristics in MESA (Race/Ethnicity: white (W), black (B), and other (Othr), and Education: less than high school (-HS), high school (HS), and more than high school (+HS))}
    \begin{tabular}{l|cc}
    \hline
     Summary & Men & Women \\ \hline
    Number of Individuals & 541 & 606 \\
    Total Observations & 2319 & 2573  \\
    Age at Enrollment & 65--80 & 65--80  \\
    Exam Schedule & 1--5 & 1--5  \\
    Race/Ethnicity & 39\% W, 28\% B, 33\% Othr & 35\% W, 33\% B, 32\% Othr  \\
    Education Level & 18\% -HS, 17\% HS, 65\% +HS & 25\% $<$HS, 22\% HS, 53\% $>$HS  \\ \hline
    \end{tabular}
    \label{Demo}
\end{table}

The MESA study measured total cholesterol at every visit, enabling the analysis of cholesterol trends over time and their relationship with CVD outcomes. 
Total cholesterol represents the overall amount of cholesterol in the blood, including low-density lipoprotein (LDL), high-density lipoprotein (HDL), and very-low-density lipoprotein (VLDL) cholesterol.
Elevated total cholesterol levels are a well-established risk factor for cardiovascular disease (CVD) as they contribute to atherosclerosis, the buildup of plaque in the arterial walls. 

In MESA, cholesterol-lowering medication use, such as statins, was recorded at each in-person exam, where participants reported their medication status. This means that information on medication use was only available at the time of each exam, and any changes in medication status occurring between exams were not directly observed. Incorporating medication use into the analysis allows for the evaluation of treatment effects, including changes in cholesterol trajectories and their potential to mitigate CVD risk. Fully observed medication status was reported for 94\% of individuals. Table~\ref{medpathern} summarizes patterns of medication use across the study, including those with missing information. The category \emph{Never on Medication} includes participants who consistently reported not taking medication throughout the study, meaning all recorded values were 0 or missing (NA). \emph{Always on Medication} includes participants who were on medication at every recorded exam, with all values being 1 or missing (NA). \emph{On and Off Medication} represents individuals who reported being on medication at some exams and off at others, meaning their data contained a mix of 0s and 1s, with or without missing values. 
\begin{table}[!t]
    \centering
    \caption{Medication use patterns in MESA}
    \begin{tabular}{l|c|c}
    \hline
    Status & Men (\%) & Women (\%) \\ \hline
    Never on Medication & 46\% & 50\% \\  
    Always on Medication & 16\% & 15\% \\  
    On and Off Medication & 38\% & 35\% \\        
    \hline
    \end{tabular}
    \label{medpathern}
\end{table}

The longitudinal nature of the MESA data is a key strength for understanding temporal patterns in CVD risk factors and medication use. The availability of serial measurements of total cholesterol, combined with repeated records of medication use, provides an opportunity to explore the interactions between these variables over time. Our model leverages the richness of the MESA data to address questions about the timing and effectiveness of interventions, accounting for missing data and the complexity of longitudinal medication patterns.

CVD events in MESA were systematically collected and adjudicated through a combination of participant follow-ups, medical record reviews, and national databases. Events were defined as the first occurrence of cardiovascular death, nonfatal myocardial infarction, stroke, transient ischemic attack, heart failure, or other cardiovascular death. Diagnoses were confirmed by physician adjudicators using pre-specified criteria, incorporating clinical symptoms, imaging, biomarkers, and medical records. In this manuscript, a CVD event specifically refers to CVD-related death.

\section{Model Specification} \label{sec3}
In this section, we introduce our model. Let \( y_{i}(a_{ij}) \) denote the risk factor level for individual \( i \) at age \( a_{ij} \), and \( m_{i}(a_{ij}) \) represent the corresponding medication status. Here, \( i = 1, \ldots, n \) indexes the individuals, and \( j = 1, \ldots, J_{i} \) represents the sequence of observed visits for individual \( i \). The switch time, \( s_{ik} \), indicates the age at which individual \( i \) undergoes their \( k \)th possible transition in medication use. It occurs in the interval \( a_{i,j-1} < s_{ik} \leq a_{ij} \), where \( k < J_{i}\) indexes the sequence of medication switches. More detail is provided in Section~\ref{secST}. \( T_i \) refers to the event time for the \( i \)-th individual.
 The structure of the joint model of risk factors, medication use, and Time-to-Event (JM-RMT) can be expressed as
\begin{eqnarray}
    y_{i}(a_{ij}) &\mid& m_{i}(a_{ij}) , \mu_{i}(a_{ij}) \notag \\
    m_{i}(a_{ij}) &\mid& m_{i}(a_{i,j-1}) , g^{\mu}(\mu_{i}(\ell); \ell\leq a_{i,j-1}) \notag \\
    s_{ik} &\mid& m_{i}(a_{i,j-1}) , m_{i}(a_{ij}) \notag \\
    T_{i} &\mid& g^{\mu}\left(\mu_{i}(\ell); \ell\leq a_{iJ_{i}}\right) , g^{m}(m_{i}(\ell); \ell\leq a_{iJ_{i}}) \notag
\end{eqnarray}
where \( g^{m}(m_{i}(\ell); \ell \leq a_{iJ_{i}}) \) and \( g^{\mu}(\mu_{i}(\ell); \ell \leq a_{iJ_{i}}) \) represent a medication feature and the risk factor feature, respectively, defined using the medication history \( m_{i}(\ell) \) and the true risk factor trajectory \( \mu_{i}(\ell) \) up to the last observed visit \( a_{iJ_{i}} \). These features incorporate all relevant information, including potential switch times between visits, to summarize the cumulative effects of medication use and risk factor progression.

Risk factors are modeled as a function of current medication status, as risk factor levels are assumed to be directly influenced by present medication use, while medication is modeled as a function of both medication history and risk factor history to capture its cumulative and dynamic nature. The hazard at age \( a_{ij} \) depends on the cumulative history of both risk factors and medication, as well as the timing of any switches. 


\subsection{Risk Factor Model}
Let \( y_{i}(a_{ij}) \) and \( m_{i}(a_{ij}) \) denote the risk factor level and medication status, respectively, of individual \( i \) at age \( a_{ij} \), where \( i = 1, \ldots, n \) and \( j = 1, \ldots, J_{i} \). We define the risk factor model as
\begin{equation}
y_{i}(a_{ij}) = \mu_{i}(a_{ij} \mid \boldsymbol{X}_{i}, \boldsymbol{b}_{i}) + \epsilon_{i}(a_{ij}),
\end{equation}
with 
\begin{equation} \mu_{i}(a_{ij} \mid \boldsymbol{X}_{i}, \boldsymbol{b}_{i}) = \mu^{-M}_{i}(a_{ij} \mid  \boldsymbol{X}_{i}, \boldsymbol{b}_{i}) + \beta_{M} a_{ij} \times m_{i}(a_{ij}) \label{mumodel}
\end{equation}
Here, $\mu^{-M}_{i}(a_{ij} \mid  \boldsymbol{X}_{i}, \boldsymbol{b}_{i})$ represents the trajectory of being off medication, and is a function of time-constant or baseline covariates $\boldsymbol{X}_{i}$ and random effects $\boldsymbol{b}_{i}$. Random intercepts and random slopes for age \( a_{ij} \) are included in the model, represented as \( \boldsymbol{b}_{i} = (b_{i0}, b_{i1}) \), where \( \boldsymbol{b}_{i} \sim \text{Normal}(\boldsymbol{0}, \boldsymbol{\Sigma}) \). The error term is denoted by $\epsilon_{i}(a_{ij})$. We assume \( \epsilon_{i}(a_{ij}) \sim SN(0, \omega, \nu) \), where \( SN \) denotes the skew normal distribution, with the scale parameter \( \omega \) and the skewness parameter \( \nu \).

\begin{figure}[!t]
    \centering
    \includegraphics[width=0.75\textwidth]{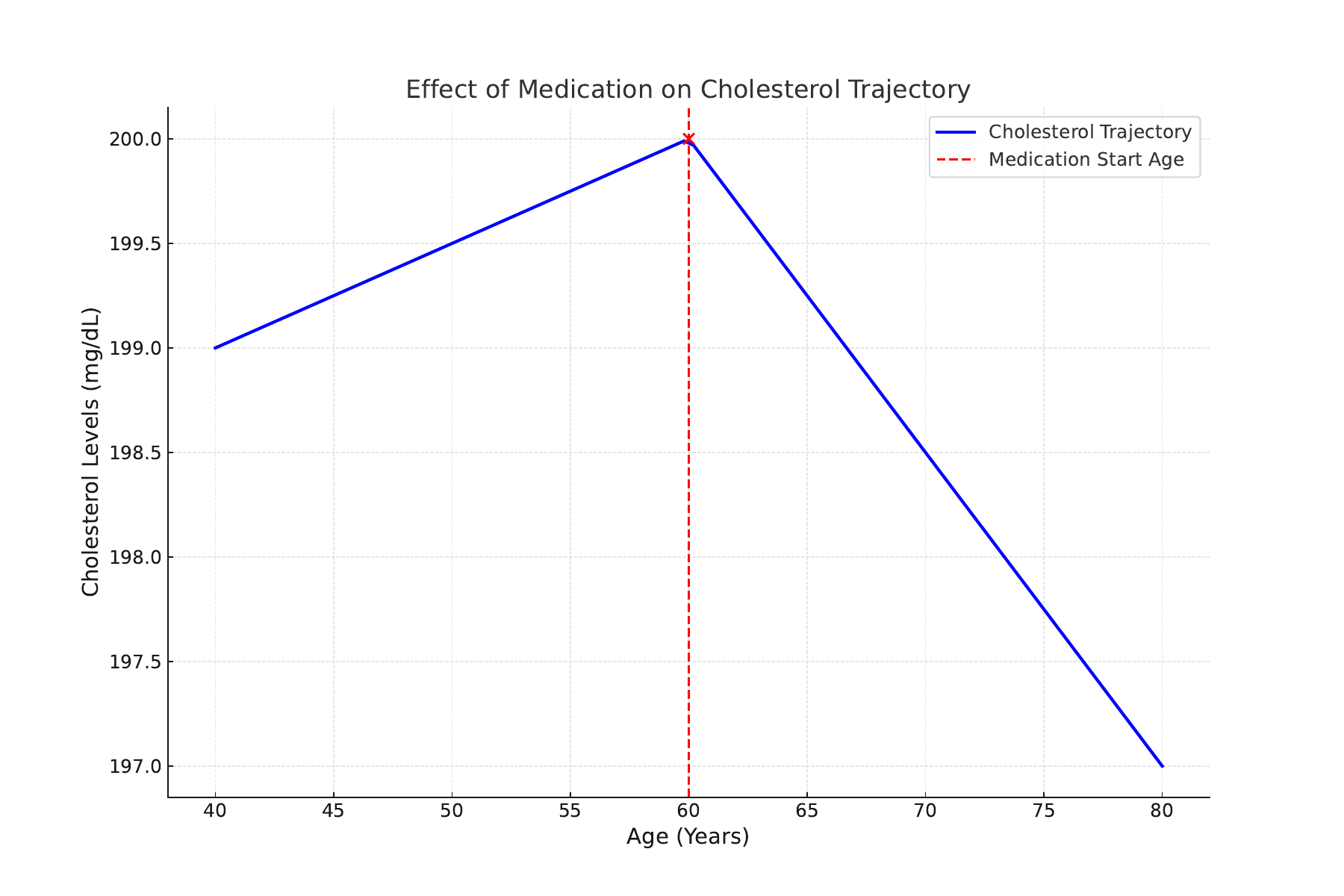}
    \caption{Illustration of the effect of medication initiation at age 60 on total cholesterol}
    \label{medchange}
\end{figure}

Starting medication often results in significant changes in the trajectory of risk factor levels at subsequent ages, introducing an interaction between age and medication status; see Figure \ref{medchange}. This interaction is captured by \( \beta_M \) in~\eqref{mumodel}, which represents the effect of medication on the rate of change in the risk factor trajectory. For individuals who have not yet started medication, it is essential to model their trajectories prior to initiation separately. This approach allows us to summarize the trends in risk factors up to the point of starting medication and then analyze the changes from the initiation of medication to the current age.

Let $\overline{m}_{i}(a_{ij})$ be the medication history. We assume
\begin{equation}
f(y_{i}(a_{ij}) \mid \overline{m}_{i}(a_{ij}), \boldsymbol{b}_{i}) = f(y_{i}(a_{ij}) \mid  m_{i}(a_{ij}), \boldsymbol{b}_{i}) \label{eq3}
\end{equation}
that is, given random effects, the current risk factor is independent of the previous medication status, and only the current medication status matters.

We denote the risk factor features by $g^{\mu}(\mu_{ i}(\ell);\ell\leq a_{ij})$, where $\mu_{ i}(\ell)$ is the true risk factor trajectory (or mean trajectory), possibly containing past information up to age $a_{ij}$. There are various choices for the risk factor feature \citep{rizopoulos2011bayesian, crowther2013adjusting, hickey2016joint}.
Here, we adopt the current value feature 
 \[
g^{\mu}(\mu_{ i}(\ell);\ell\leq a_{ij}) = \mu_{i}(a_{ij}),
\]
which assumes that the associated medication and hazard models at age $a_{ij}$ are directly related to the value of the linear predictor from the longitudinal submodel, evaluated at the same age $a_{ij}$; see Sections \ref{med} and \ref{haz} for further details. However, for the sake of generality, we continue to denote risk factor features by \( g^{\mu}(\mu_i(\ell), \ell \leq a_{ij}) \) in the remainder of this manuscript.


\subsection{Medication Model} \label{med}
Let $g^{\mu}(\mu_{i}(\ell); \ell \leq a_{i,j-1})$ and $\overline{m}_{i}(a_{ij})$ denote the risk factor feature up to age $a_{i,j-1}$ (the age at the previous observed exam) and the medication history up to age $a_{ij}$, respectively. We assume
\begin{equation}
f\left(m_{i}(a_{ij}) \mid g^{\mu}(\mu_{i}(\ell); \ell \leq a_{i,j-1}), \overline{m}_{i}(a_{ij})\right) = 
f\left(m_{i}(a_{ij}) \mid g^{\mu}(\mu_{i}(\ell); \ell \leq a_{i,j-1}), m_{i}(a_{i,j-1})\right), 
\label{eqmed1}
\end{equation}
such that the current medication status $m_{i}(a_{ij})$ depends on both the individual’s risk factor history and their previous medication status $m_{i}(a_{i,j-1})$, but not on the medication history prior to $a_{i,j-1}$. 

The medication model is specified as
\begin{eqnarray}
   && \text{logit}\left[Pr (m_{i}(a_{ij}) = 1 \mid m_{i}(a_{i,j-1}) = m, g^{\mu}(\mu_{i}(\ell); \ell \leq a_{i,j-1})) \right] \notag \\
   &&= \gamma_{0}^{(m)}(a_{ij}) + \gamma_{1}^{(m)}(a_{ij} - a_{i,j-1}) + \gamma_{2}^{(m)}g^{\mu}(\mu_{i}(\ell); \ell \leq a_{i,j-1}), 
   \label{medmodel}
\end{eqnarray}
where $m = 0,1$, and $\mu_{i}(\ell)$ denotes the true risk factor value at age $\ell$, and $g^{\mu}(\mu_{i}(\ell); \ell \leq a_{i,j-1})$ refers to the risk factor feature up to age $a_{i,j-1}$. In (\ref{medmodel}), $\gamma_{0}^{(m)}(\cdot)$ is a smooth function of age $a_{ij}$, representing the baseline medication probability. The term $\gamma_{1}^{(m)}(a_{ij} - a_{i,j-1})$ is a smooth function of the lagged age $a_{ij} - a_{i,j-1}$, capturing the temporal stability/dependence of medication effects over time. Finally, $\gamma_{2}^{(m)}$ represents the effect of the mean risk factor feature $g^{\mu}$ up to age $a_{i,j-1}$.

In this study, age is recorded as the integer age at the last birthday (e.g., Age = $\lfloor \text{real age} \rfloor$). The lagged age effect $\gamma_{1}(a_{ij} - a_{i,j-1})$ is modeled as
\[
\gamma_{1}(a_{ij} - a_{i,j-1}) = \exp\left(-\lambda \left| a_{ij} - a_{i,j-1} \right|\right),
\]
which monotonically decreases towards zero as the lag increases. This formulation ensures that, for longer lags, the probability of medication at age $a_{ij}$ depends primarily on $\gamma_{0}^{(m)}$ and $\gamma_{2}^{(m)}$, thereby mitigating concerns about extrapolation. In practical terms, if a significant gap exists between consecutive ages, the model relies on the baseline and risk factor feature effects, and not medication status at the (distant) prior exam. This approach ensures that medication effects are not extrapolated too far beyond observed data points. However, it also means that if medication use changed within the gap but was unrecorded, the model cannot explicitly capture these transitions.

The medication feature, denoted as \( g^{M}(m_{i}(\ell); \ell \leq a_{ij}) \), can take on various possible representations. These include the medication status of the previous year, current medication status, the number of years on medication (cumulative medication use), or a function reflecting the number of medication switches over the past years. Each of these formulations captures different aspects of the influence of medication on the hazard model, and the choice depends on the specific context.
We specify the medication feature as the number of years on medication, which depends on both medication history and switch time. Switch time refers to the age at which a subject transitions on or off medication and is required to compute the medication feature.


\subsection{Modeling Medication Use Between Visits} \label{secST}
When calculating a cumulative medication use feature, it is important to account for the uncertainty in medication status between visits. Since medication status is only measured at the time of the exam, and MESA exams are at least two years apart, there is uncertainty in medication status between exams that must be accounted for. Three primary sources of uncertainty can arise in this context. First, uncertainty can occur in adjacent visits. When \(m_{i}(a_{i,j-1}) = m_{i}(a_{ij})\), we assume there is no switch in medication status between the two visits, given the short 2-year window between visits (i.e., do not allow two switches in this window). However, when \(m_{i}(a_{i,j-1}) \neq m_{i}(a_{ij})\), this indicates that a switch occurred at some point between the two study visits, though the exact timing of the switch is unknown.

A second source of uncertainty arises from gaps between visits due to a missed study visit or missing medication status. In the case of a gap involving one visit, if \(m_{i}(a_{i,j-2}) = m_{i}(a_{ij})\), we cannot conclusively assume that a change in medication status did not occur during the gap. Conversely, if \(m_{i}(a_{i,j-2}) \neq m_{i}(a_{ij})\), it becomes possible for more than one switch to have occurred within the interval. 

\begin{figure}[!t]
    \centering
    \includegraphics[width=0.9\textwidth]{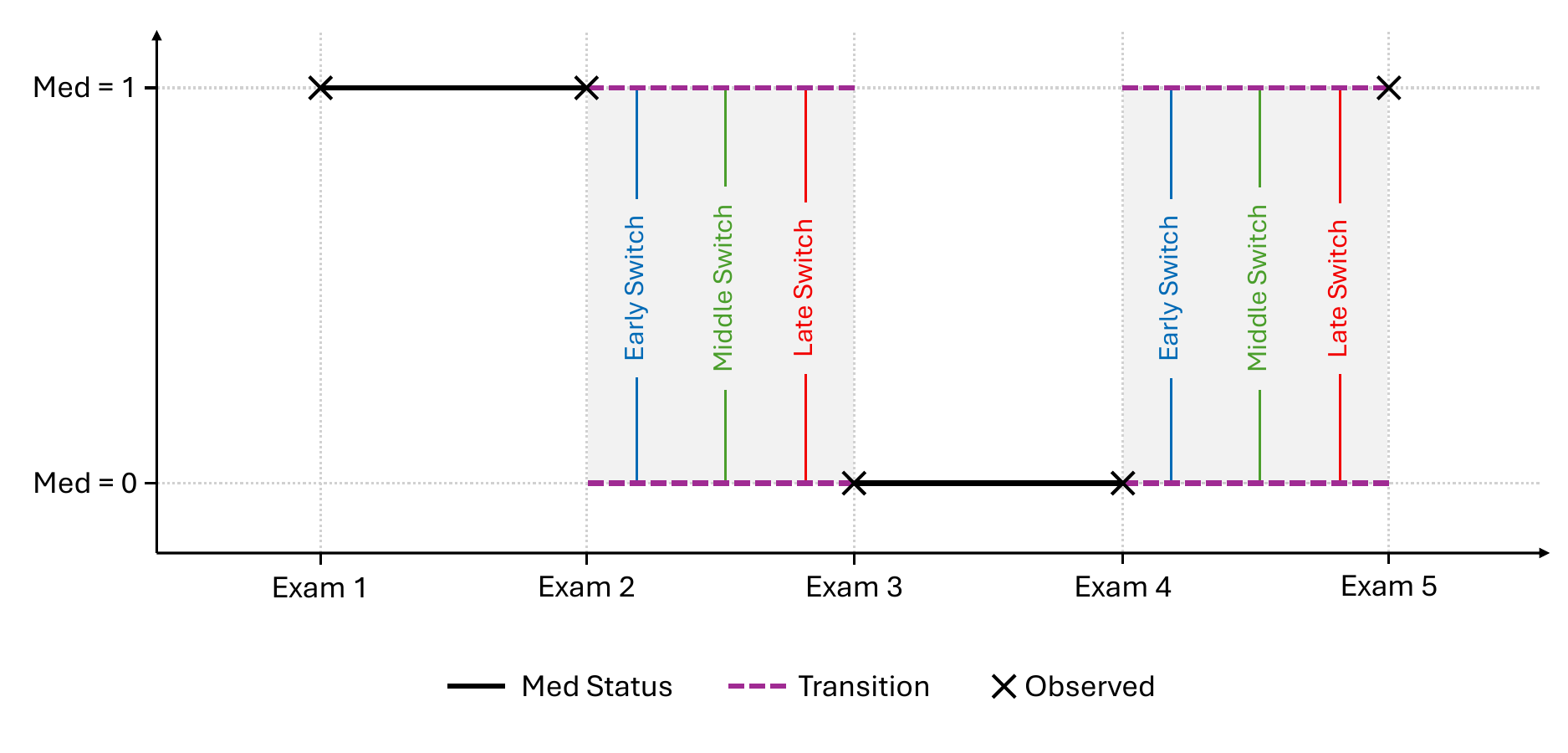}
    \caption{Illustration of possible medication status transitions across exams for a participant who switched off medication between Exams 2 and 3, then resumed medication between Exams 4 and 5.}
    \label{switching}
\end{figure}

Figure \ref{switching} illustrates the potential medication status transitions across exams, showcasing early, middle, and late switch scenarios between consecutive exams and their corresponding trajectories, highlighting the dynamic nature of medication usage patterns over time. In Figure \ref{switching}, the black horizontal lines represent the binary medication status (Med = 0 or 1) at each exam, while the transitions between exams are shown in purple to mark the moments of change. The annotations ‘Early Switch,’ ‘Middle Switch,’ and ‘Late Switch’ emphasize the timing of transitions relative to the five exam periods. For instance, an early switch occurs shortly after the second exam, whereas a late switch occurs closer to the fifth exam. These patterns underscore the variability in medication initiation and cessation, reflecting differences in individual treatment plans or adherence behaviors.

For the $i^{th}$ individual, when \(m_{i}(a_{i,j-1}) \neq m_{i}(a_{ij})\), let \( s_{ik} \in \{ a_{i,j-1}+1, \ldots, a_{ij} \}\) represent the possible integer ages between \( a_{i,j-1} \) and \( a_{ij} \) where a switch could occur. We assume that the probabilities of switch times are proportional to the probabilities from the medication model in (\ref{medmodel}),
\begin{equation}
\Gamma_{i} (s_{ik}) = \gamma_{0}^{(m)}(s_{ik}) + \gamma_{1}^{(m)}(s_{ik}-a_{i,j-1}) + \gamma_{2}^{(m)} g^{\mu}(\mu_{i}(\ell); \ell \leq a_{i,j-1}),
\label{stmodel}
\end{equation}
and specify the distribution of the switch time at integer ages via normalization as
\begin{equation}
f \left( s_{ik} \mid m_{i}(a_{i,j-1}), m_{i}(a_{ij}) \right) = \frac{\text{logit}^{-1} \left( \Gamma_{i} (a_{ik}) \right)}{\sum_{k \in \{ a_{i,j-1}+1, \ldots, a_{ij} \}} \text{logit}^{-1} \left( \Gamma_{i} (a_{ik}) \right)}. \label{stmodeldensity}
\end{equation}
\( s_{ik} \) follows a multinomial distribution. The observed data provide no direct information about the exact switch time, but this approach allows us to account for uncertainty in the switch location exploiting the medication model.

If we had complete annual data on medication status or continuous data such as refill patterns, we would not need to model them to calculate the medication feature; they would be treated as observed time-varying covariates in the hazard function. In the absence of annual data, even when medication status data is fully observed at every visit, modeling switch times is important because we cannot overlook the potential for switches to occur between visits. 
This is where methods like LOCF fall short. If a participant switches early, LOCF can overestimate the years on medication by a significant amount, depending on the time between the last observed measurement and the point of switch. Conversely, for participants who switch off medication later, LOCF may underestimate the true duration off medication. These biases highlight the need for methods that account for the timing of medication switches, as LOCF can lead to inaccurate estimates of medication exposure.
Note that if observations were recorded annually, with \( (a_{ij} - a_{i,j-1}) = 1 \), the function $\gamma_{1}(\cdot)$ in (\ref{stmodel}) simplifies to \( e^{-\lambda} \).
For individual \(i\) with \(J_{i}\) visits, a maximum of \(J_{i} - 1\) switches can occur during the follow-up period. 

When medication status is missing at a visit, both switching and missingness issues arise. Under an assumption of ignorable missingness, this can be accommodated using data augmentation. For instance, for a subject with four visits and missing medication status at Exam 2, we impute the missing medication status with probability proportional to the product of \( Pr\left(m_{i}(a_{i2}) \mid m_{i}(a_{i1})\right) \) and \( Pr\left(m_{i}(a_{i3}) \mid m_{i}(a_{i2})\right) \) from the medication model (\ref{medmodel}).

\begin{figure}[!t]
    \centering
    \includegraphics[width=1\textwidth]{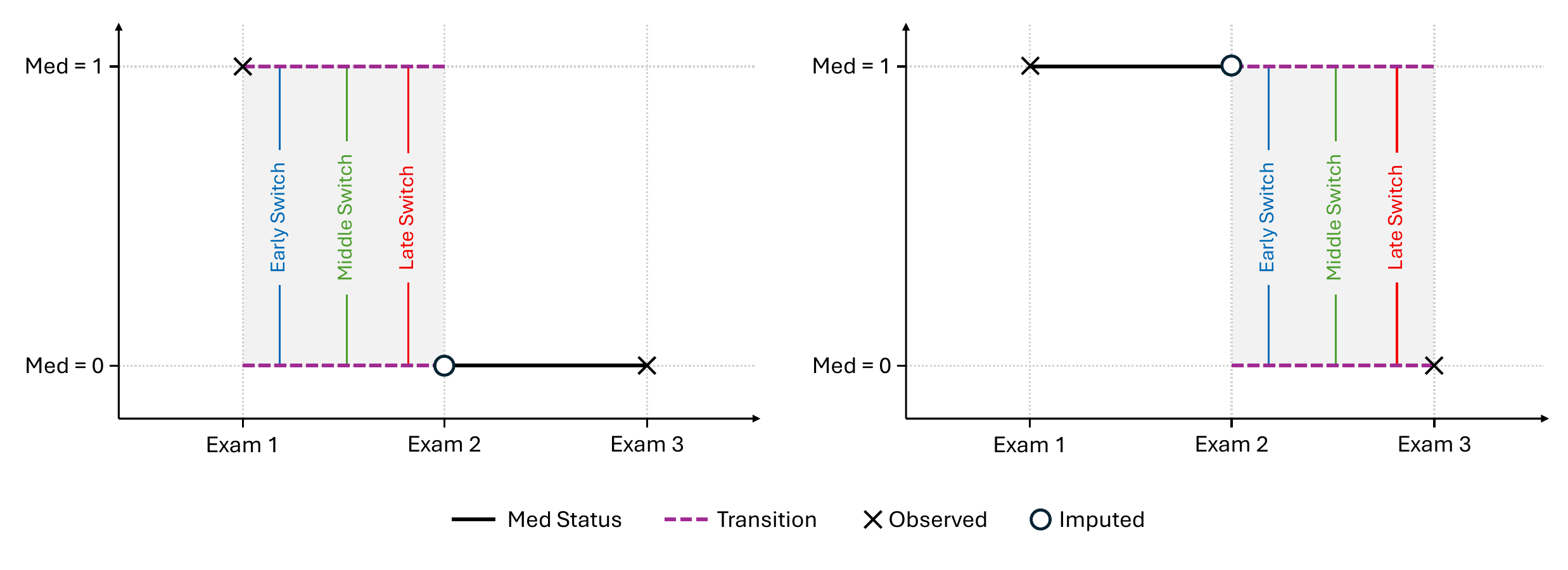}
    \caption{Possible medication status scenarios for a participant with missing medication status at Exam 2 who switched off medication by Exam 3.}
    \label{missingmed}
\end{figure}

Figure \ref{missingmed} illustrates two possible scenarios for imputing missing medication data when medication status is missing at Exam 2. The model imputes whether the individual remained on medication or switched off. In the left panel, if the imputed value (indicated by a circle) is off medication at Exam 2, the switch is assumed to occur during the gap between Exams 1 and 2. In the right panel, the missing status at Exam 2 is imputed as being on medication, so that switching off occurs between Exams 2 and 3. The vertical lines represent the likely timing of transitions, categorized as early, middle, or late switches. These scenarios highlight the importance of accounting for both missing data and transitions to accurately capture the uncertainty and its impact on the analysis of medication patterns and cumulative medication use.

Once the missing medication statuses are imputed using data augmentation, we can calculate the medication feature to evaluate its impact on survival and its influence on risk factor trajectories. This approach enables us to answer critical questions, such as how an individual’s medication history affects their survival probabilities and how medication impacts underlying risk factors over time.
Although \( s_{ik} \) does not impact the medication model, it is needed to compute the medication feature in the hazard model.

\subsection{Hazard Model} \label{haz}
We denote by \( T_i \) the true event time for the \( i \)th subject, \( T^*_i \) the observed event time, defined as the minimum of the potential censoring time \( C_i \) and \( T_i \); and by \( \delta_i = I(T_i \leq C_i) \) the event indicator.
 For the \( i \)th participant, \( i = 1, \ldots, n \), at age \( a_{ij} \), we define the hazard model for terminal events as 
\begin{eqnarray}
  && h_{i}\left(t\mid \boldsymbol{X}_{i}, g^{\mu}\left(\mu_{i}(\ell); \ell \leq \min(t,a_{iJ_{i}})\right), g^{M}\left(m_{i}(\ell); \ell \leq \min(t,a_{iJ_{i}})\right)\right) = \notag \\
  && \quad h_{i0}(t) \exp\{\boldsymbol{X}_{i}\boldsymbol{\beta}_h + \lambda^{\mu} u^{\mu}(t-a_{iJ_{i}})^{I\{t>a_{iJ_{i}}\}}g^{\mu}\left(\mu_{i}(\ell); \ell \leq \min(t,a_{iJ_{i}})\right) +\notag \\
  && \quad \quad \lambda^{M} u^{m}(t-a_{iJ_{i}})^{I\{t>a_{iJ_{i}}\}}g^{M}\left(m_{i}(\ell); \ell \leq \min(t,a_{iJ_{i}})\right)\},
\end{eqnarray}
where \( \boldsymbol{X}_{i} \) is a matrix of baseline covariates, and \( \boldsymbol{\beta}_h \) is a vector of regression coefficients. The term \( a_{iJ_{i}} \) represents the age at the last observed visit, up to which the risk factor feature \( g^{\mu}(\cdot) \) and the medication feature \( g^{M}(\cdot) \) are defined based on available data. Beyond \( a_{iJ_{i}} \), no further data are collected until the event or censoring time.
This represents the third source of uncertainty. To address this, \( u^{\mu}(\cdot) \) and \( u^{m}(\cdot) \) are smooth, increasing functions that account for the lack of feature contributions during the period between \( a_{iJ_{i}} \) and the event time \( T_i \). These functions ensure that the effects of the risk factor and medication features are appropriately adjusted and gradually diminish as time progresses without additional observations. \( h_{i0}(a_{ij}) \) is the baseline hazard, specified using B-splines as
\begin{equation}
\log h_0(a_{ij}) = \kappa_0 + \sum_{d=1}^{m} \kappa_d B_d(a_{ij}, q),
\end{equation}
where \( \kappa^\top = (\kappa_0, \kappa_1, \ldots, \kappa_m) \) are the spline coefficients, \( q \) denotes the degree of the B-spline basis functions \( B(\cdot) \), and the number of B-spline basis functions is given by \( D = \ddot{D} + q - 1 \), with \( \ddot{D} \) denoting the number of interior knots, and consequently, the number of spline coefficients $m = D $.

The survival function can be expressed as
\begin{eqnarray}
  && S_{i}\left(a_{ij}\mid \boldsymbol{X}_{i}, g^{\mu}(\mu_{i}(t); t \leq a_{ij}), g^{M}(m_{i}(t); t \leq a_{ij})\right)
    = \notag \\
    && \quad \exp\left\{-\int_{a_{i1}}^{T_{i}} h_{i}\left(s \mid \boldsymbol{X}_{i}, g^{\mu}(\mu_{i}(\ell); \ell \leq \min(s, a_{iJ_{i}})), g^{M}(m_{i}(\ell); \ell \leq \min(s, a_{iJ_{i}}))\right) ds\right\}.
\end{eqnarray}
Individual \( i \) contributes no data until \( a_{i1} \), and the lower bound of the integral can be the age at baseline, i.e., \( \{a_{i1}\}_{i=1,\ldots,n} \). We use Gauss-Kronrod quadrature with
Q nodes to approximate the survival function, ensuring precise numerical integration over age intervals. Further details on this numerical approximation method can be found in Supplementary Materials (\hyperref[S.1.]{S.1}).


The likelihood component for the events is obtained by multiplying the hazard with the survival function,
{
\thinmuskip=0.1mu
\thickmuskip=0.1mu
\begin{eqnarray}
&& P\left(T_i, \delta_i \mid \boldsymbol{X}_{i}, g^{\mu}(\mu_{i}(t); t \leq T_{i}), g^{M}(m_{i}(t); t \leq T_{i})\right) = \notag \\
&& \quad h_{i}\left(T_{i} \mid \boldsymbol{X}_{i}, g^{\mu}(\mu_{i}(t); t \leq \min(T_{i}, a_{iJ_{i}})), g^{M}(m_{i}(t); t \leq \min(T_{i}, a_{iJ_{i}}))\right)^{\delta_i} \notag \\
&& \quad \times ~ S_{i}\left(T_{i} \mid \boldsymbol{X}_{i}, g^{\mu}(\mu_{i}(t); t \leq \min(T_{i}, a_{iJ_{i}})), g^{M}(m_{i}(t); t \leq \min(T_{i}, a_{iJ_{i}}))\right),
\label{eventlike}
\end{eqnarray}
}%
where \( T_i \) and \( \delta_i \) represent the observed event time and event indicator for the \( i \)th participant, \( i = 1, \ldots, n \), respectively.

\subsection{Likelihood}
Let  \( \mathcal{J}_{i} = \{j=1,\ldots,J_{i} \mid m_{i}(a_{i,j-1}) \neq m_{i}(a_{ij}) \} \) be the set of all consecutive visits where a switch occurred, and \( \mathcal{C}_{ij}=  \{ a_{i,j-1}+1, \ldots, a_{ij} \} \) be the set of all the possible integer ages between \( a_{i,j-1} \) and \( a_{ij} \) where a switch could occur. When there is no missing medication status, the observed data likelihood  for subject $i$, which averages over switch times, is
\begin{eqnarray}
 L^{obs}_{i} &=& 
\sum_{j \in \mathcal{J}_{j}} ~    
\sum_{s_{ij}\in \mathcal{C}_{ij}}
\text{\Huge [ }\prod_{j=1}^{J_{i}}  f\left(y_{i}(a_{ij})\mid  m_{i}(a_{ij}), b_{i}\right)  \notag \\
&& \quad  \times \prod_{j=2}^{J_{i}} f\left(m_{i}(a_{ij})\mid  m_{i}(a_{i,j-1}) , g^{\mu}(\mu_{i}(\ell); \ell\leq a_{i,j-1})\right) \notag  \\ 
&& \quad  \times\prod_{j=2}^{J_{i}}f\left(s_{i,j-1} \mid m_{i}(a_{i,j-1}) , m_{i}(a_{ij})\right)^{I\{m_{i}(a_{i,j-1}) \neq m_{i}(a_{ij})\}}  \notag \\ 
&& \quad  \times f\left(T_i, \delta_{i}\mid g^{\mu}\left(\mu_{i}(\ell); \ell\leq a_{iJ_{i}})\right) , g^{m}(m_{i}(\ell); \ell\leq a_{iJ_{i}})\right) \text{\Huge ] },   \notag
\end{eqnarray}
where the last term is given in (\ref{eventlike}).

The observed data likelihood differs when medication status is missing.
For example, when \( m_{i}(a_{i2}) \) is missing, the observed data likelihood is
\begin{eqnarray}
 L^{obs}_{i} &=&
\sum_{m_{i}(a_{i2})=0}^{1} ~ 
\sum_{j \in \mathcal{J}_{j}} ~    
\sum_{s_{ij}\in \mathcal{C}_{ij}}
\text{\Huge [ }\prod_{j=1}^{J_{i}}  f\left(y_{i}(a_{ij})\mid  m_{i}(a_{ij}), b_{i}\right)  \notag \\
&& \quad  \times \prod_{j=2}^{J_{i}} f\left(m_{i}(a_{ij})\mid  m_{i}(a_{i,j-1}) , g^{\mu}(\mu_{i}(\ell); \ell\leq a_{i,j-1})\right) \notag  \\ 
&& \quad  \times\prod_{j=2}^{J_{i}}f\left(s_{i,j-1} \mid m_{i}(a_{i,j-1}) , m_{i}(a_{ij})\right)^{I\{m_{i}(a_{i,j-1}) \neq m_{i}(a_{ij})\}}  \notag \\ 
&& \quad \times f\left(T_i, \delta_{i}\mid g^{\mu}\left(\mu_{i}(\ell); \ell\leq a_{iJ_{i}})\right) , g^{m}(m_{i}(\ell); \ell\leq a_{iJ_{i}})\right) \text{\Huge ] } .  \notag
\end{eqnarray}

\sloppy
The observed data likelihood incorporates contributions from the longitudinal submodel, the medication model, the switch time, and the time-to-event model. 
Specifically, \( f\left(y_{i}(a_{ij})\mid  m_{i}(a_{ij}), b_{i}\right) \) captures the longitudinal outcomes as a function of current medication status and random effects, \( f\left(m_{i}(a_{ij})\mid  m_{i}(a_{i,j-1}) , g^{\mu}(\mu_{i}(\ell); \ell\leq a_{i,j-1})\right) \) denotes the probability of current medication status given the risk factor history and the previous medication status. The term \( f\left(s_{i,j-1} \mid m_{i}(a_{i,j-1}) , m_{i}(a_{ij})\right)^{I\{m_{i}(a_{i,j-1}) \neq m_{i}(a_{ij})\}} \) is the distribution of switch times when a change in medication occurs. If no switch takes place, this term is omitted by the indicator function $I\{m_{i}(a_{i,j-1}) \neq m_{i}(a_{ij})\}$. Finally, \( f\left(T_i, \delta_{i}\mid g^{\mu}\left(\mu_{i}(\ell); \ell\leq a_{iJ_{i}})\right), g^{m}(m_{i}(\ell); \ell\leq a_{iJ_{i}})\right) \) reflects the contributions from the time-to-event model, incorporating both risk factor and medication history up to the last observed visit.


\subsection{Inference}
We use Bayesian inference to estimate the parameters of our joint model. Posterior sampling is conducted using the \texttt{Nimble} package for R \citep{nimble-software:2024}. Details of the prior specifications are given in Supplementary Materials (\hyperref[S.2.]{S.2}).


\section{Application to Longitudinal Cholesterol Outcomes in MESA} \label{sec4}
This section summarizes the results of fitting JM-RMT to the MESA data. We first provide further details on the model specification and the MESA data. Our model incorporates key demographic covariates to capture variations in risk trajectories and event hazards. 
In the longitudinal model, demographic covariates such as education and race are included in \( \mu^{-M}_{i}(a_{ij} \mid \boldsymbol{X}_{i}, \boldsymbol{b}_{i}) \). These covariates interact with age, allowing us to evaluate how education level and race influence risk factor trajectories over time. Additionally, a random intercept and a random slope for age are incorporated to capture individual variability in baseline risk and age-related trajectories.
Education level is represented as a categorical variable with three groups; less than high school (-HS), high school (HS), and more than high school (+HS). Race is included as a binary variable, distinguishing between Black and non-Black individuals. These covariates are also included in the hazard model. 

\subsection{Posterior Sampling}
We use four independent chains to assess convergence. Each chain includes 150k samples, with the first 50k discarded.
Convergence diagnostics, including visual inspection of trace plots and the Gelman-Rubin statistic (\( \hat{R} \)), confirmed that all chains adequately converged. Posterior distributions for all model parameters were summarized using means, standard deviations, and credible intervals based on the retained MCMC samples. The model's structure accommodates ignorable missing values using data augmentation.
Posterior estimates for the covariate effects are presented in Supplementary Materials (\hyperref[S.3.]{S.3}).

\subsection{Medication Model Specification}
In the medication model in~(\ref{medmodel}), \( \gamma_{0}^{(m)}(a_{ij}) \) is specified as a linear function of age, capturing the baseline effect of age on the probability of medication use; \( \gamma_0^{(0)} = \alpha_1 + \alpha_3 a_{ij} \) represents the effect of age when the previous medication status is \( m = 0 \), while \( \gamma_0^{(1)} = (\alpha_1 + \alpha_2) + (\alpha_3 + \alpha_4) a_{ij} \) captures the effect of age when the previous medication status is \( m = 1 \). Similarly, \( \gamma_{1}^{(m)}(a_{ij} - a_{i,j-1}) \) is specified as an exponential decay function
, ensuring that the influence of the gap between visits on medication transitions diminishes gradually; \( \gamma_1^{(0)} = \alpha_5 \exp(-|a_{ij} - a_{i,j-1}|) \) when \( m = 0 \), and \( \gamma_1^{(1)} = (\alpha_5 + \alpha_6) \exp(-|a_{ij} - a_{i,j-1}|) \) when \( m = 1 \). Further details on the parametrization of the medication model are provided in Supplementary Materials (\hyperref[S.3.]{S.3}).
\begin{figure}[!t]
    \centering 
    \subfloat[Men]
    {\includegraphics[width=0.5\textwidth]{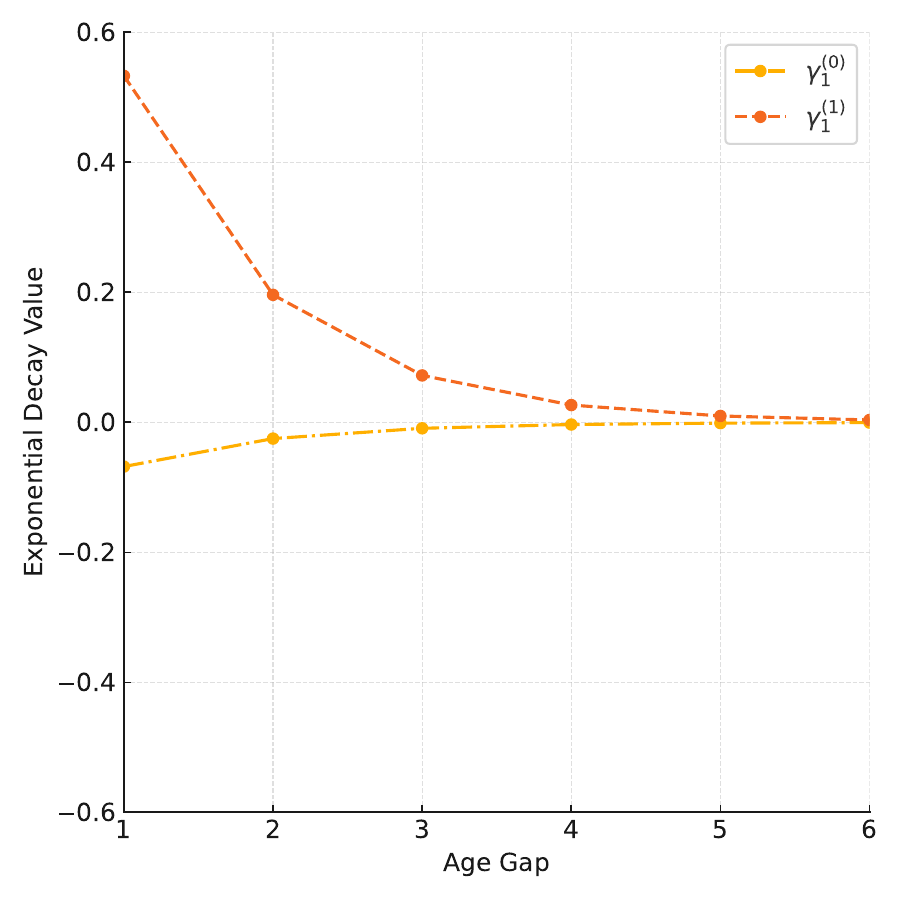}} 
    \subfloat[Women]{\includegraphics[width=0.5\textwidth]{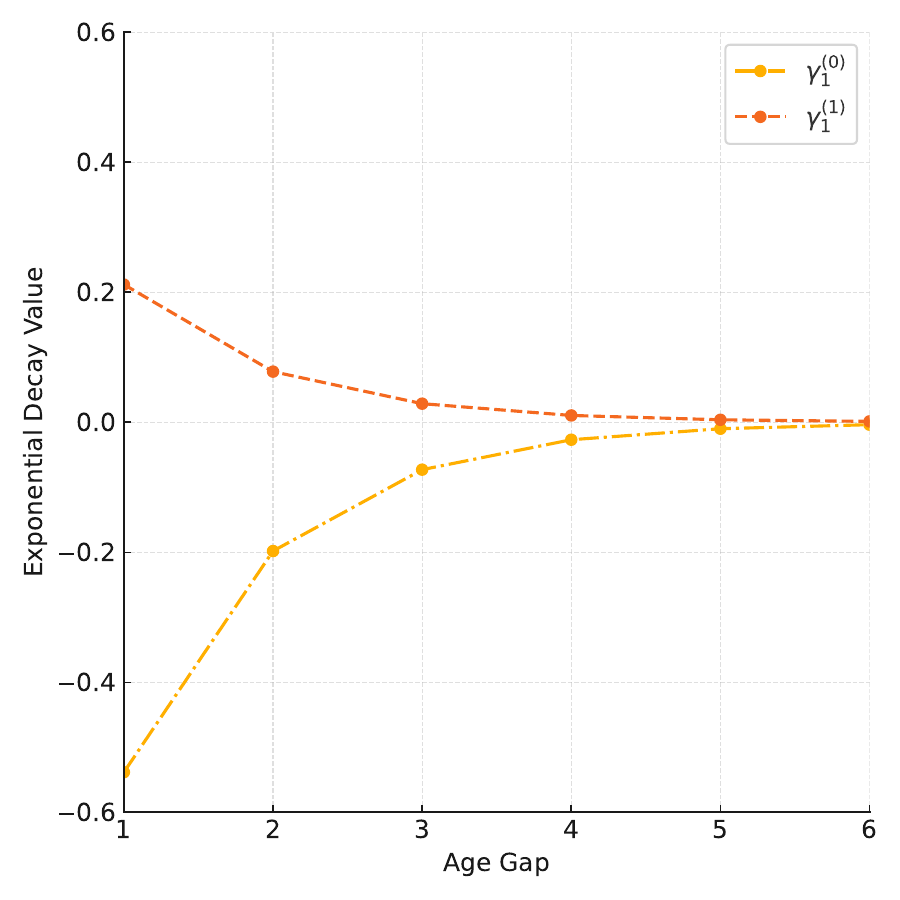}} 
    \caption{Posterior mean of the exponential decay functions for on- and off-medication conditions in (a) men and (b) women}
    \label{decay_comparison}
\end{figure}

Figure~\ref{decay_comparison} shows posterior estimates of \( \gamma_{1} \) as a function of visit gap for on- and off-medication conditions in men and women. For men, the on-medication effect (\( \gamma_{1}^{(1)} \)) demonstrates a higher initial impact and slower decay, reflecting a sustained benefit of medication, while the off-medication effect (\( \gamma_{1}^{(0)} \)) declines rapidly, indicating a swift loss of benefit upon discontinuation. In women, the on-medication effect starts lower and decays faster, suggesting a relatively diminished and shorter-lived benefit of medication, with a similarly rapid decay for off-medication effects.

These findings align with prior evidence of gender-specific differences in cholesterol-lowering responses to statins, where men generally experience greater efficacy in lowering LDL cholesterol compared to women~\citep{kostis2012meta}. However, the relative risk reduction (RRR) for CVD events is similar between men and women, and adherence to therapy is often lower among women, potentially contributing to these differences~\citep{lewey2013gender}. 

\subsection{Comparison with LOCF}
\subsubsection{Posterior estimates}
We compare the parameters estimated by JM-RMT with those derived from LOCF, which assumes that the last observed medication status remains constant until the next observation.
Furthermore, we also fit two models in which the risk factor feature was excluded from the survival model to examine its effect on the coefficient for the medication feature. This analysis allows us to disentangle the contributions of the medication feature and the risk factor feature to CVD events, providing a clearer understanding of their individual and combined roles.

\begin{table}[!t]
\centering
\caption{Posterior summaries of the coefficients of the medication feature and the risk factor feature in the survival model across methods and sex}
\begin{adjustbox}{max width=\textwidth}
\begin{tabular}{lrrrrrrrrr}
\toprule
 &  & \multicolumn{4}{c}{Men} & \multicolumn{4}{c}{Women} \\
\cmidrule(lr){3-6} \cmidrule(lr){7-10}
  Method     &  & Mean & SD & 2.5\% & 97.5\% & Mean & SD & 2.5\% & 97.5\% \\
\midrule
\emph{Hazard including $g^{\mu}$}&&&&&&&&& \\
JM-RMT & $\lambda^{M}$ & -0.688 & 0.413 & -1.634 & -0.039 & 0.047 & 0.144 & -0.307 & 0.246 \\
           & $\lambda^{\mu}$ & -0.033 & 0.008 & -0.050 & -0.021 & -0.025 & 0.007 & -0.042 & -0.014 \\
LOCF & $\lambda^{M}$ & -0.061 & 0.326 & -0.782 & 0.503 & 0.648 & 0.284 & 0.127 & 1.247 \\
& $\lambda^{\mu}$ & -0.039 & 0.010 & -0.061 & -0.023 & -0.042 & 0.014 & -0.074 & -0.020 \\
\emph{Hazard excluding $g^{\mu}$}&&&&&&&&& \\
JM-RMT  & $\lambda^{M}$ & -0.980 & 0.357 & -1.798 & -0.410 & -0.237 & 0.167 & -0.621 & 0.026 \\
LOCF & $\lambda^{M}$ & -0.719 & 0.276 & -1.350 & -0.277 & -0.159 & 0.159 & -0.516 & 0.104 \\
\bottomrule
\end{tabular}
\end{adjustbox}
\label{results1}
\end{table}
Table \ref{results1} presents the coefficients of the medication feature (years on medication) and the risk factor feature (current value) on the hazard of CVD events across methods and sexes. In the JM-RMT, years on medication (\( \lambda^{M} \)) demonstrates a significant protective effect for men, while the effect for women is smaller and not statistically significant. By contrast, the LOCF method underestimates the protective effect for men and overestimates it for women, with a significant effect in the wrong direction for women, illustrating the limitations of assuming no change in medication status between visits. The current risk factor value (\( \lambda^{\mu} \)) consistently shows a significant and protective effect in all models, with slightly stronger effects observed for men in the JM-RMT. Excluding the risk factor feature (\( g^{\mu} \)) amplifies the effect of years on medication in all cases, suggesting that part of its protective effect may be mediated through the current risk factor value. These findings emphasize the ability of JM-RMT to more accurately capture the interplay between medication use, risk factor trajectories, and CVD risk, while exposing the potential biases of naïve approaches like LOCF.

\begin{figure}[!t]
    \centering
    \subfloat[Men]{
    \includegraphics[width=0.49\textwidth]{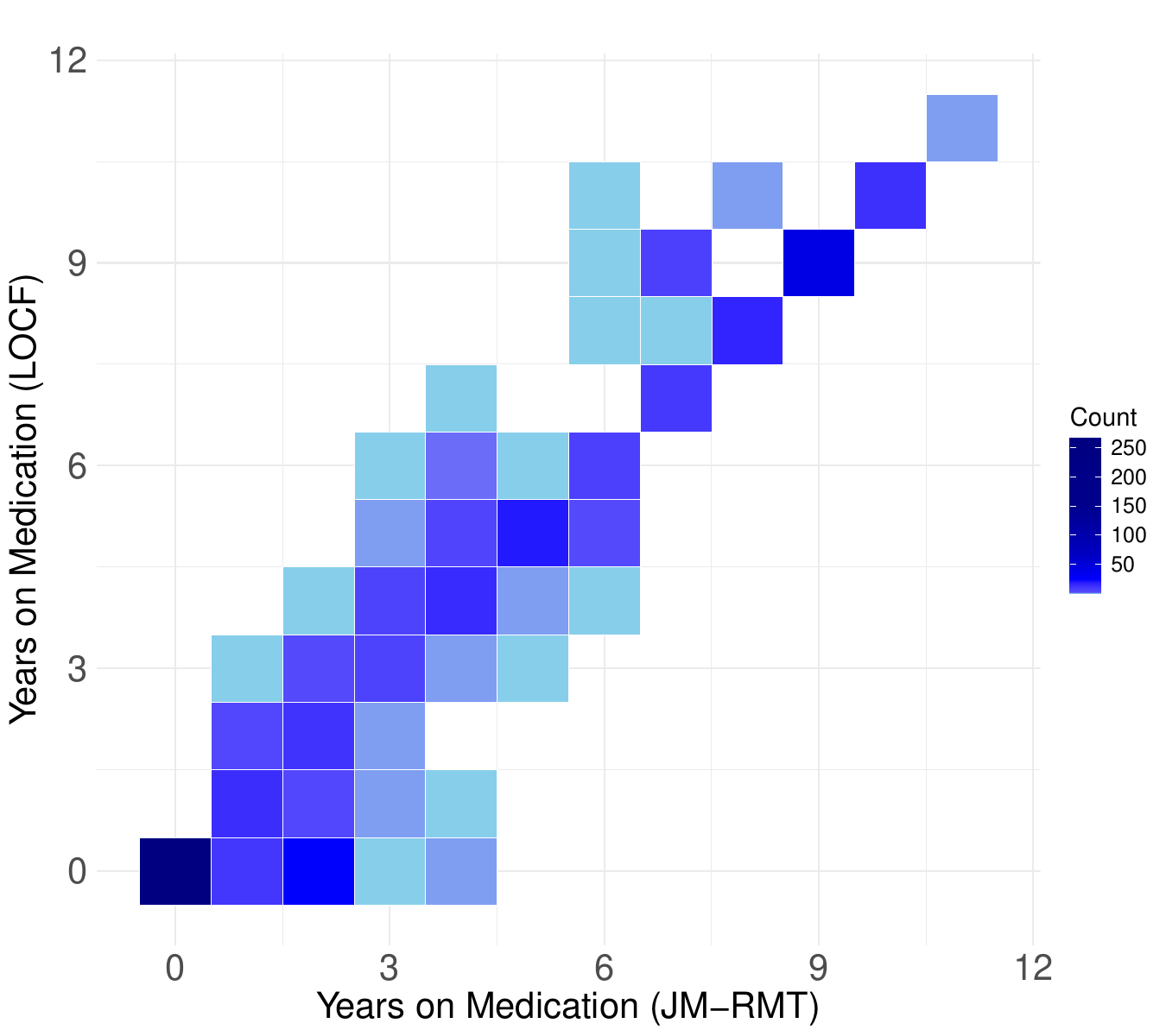}}
    \subfloat[Women]{
    \includegraphics[width=0.49\textwidth]{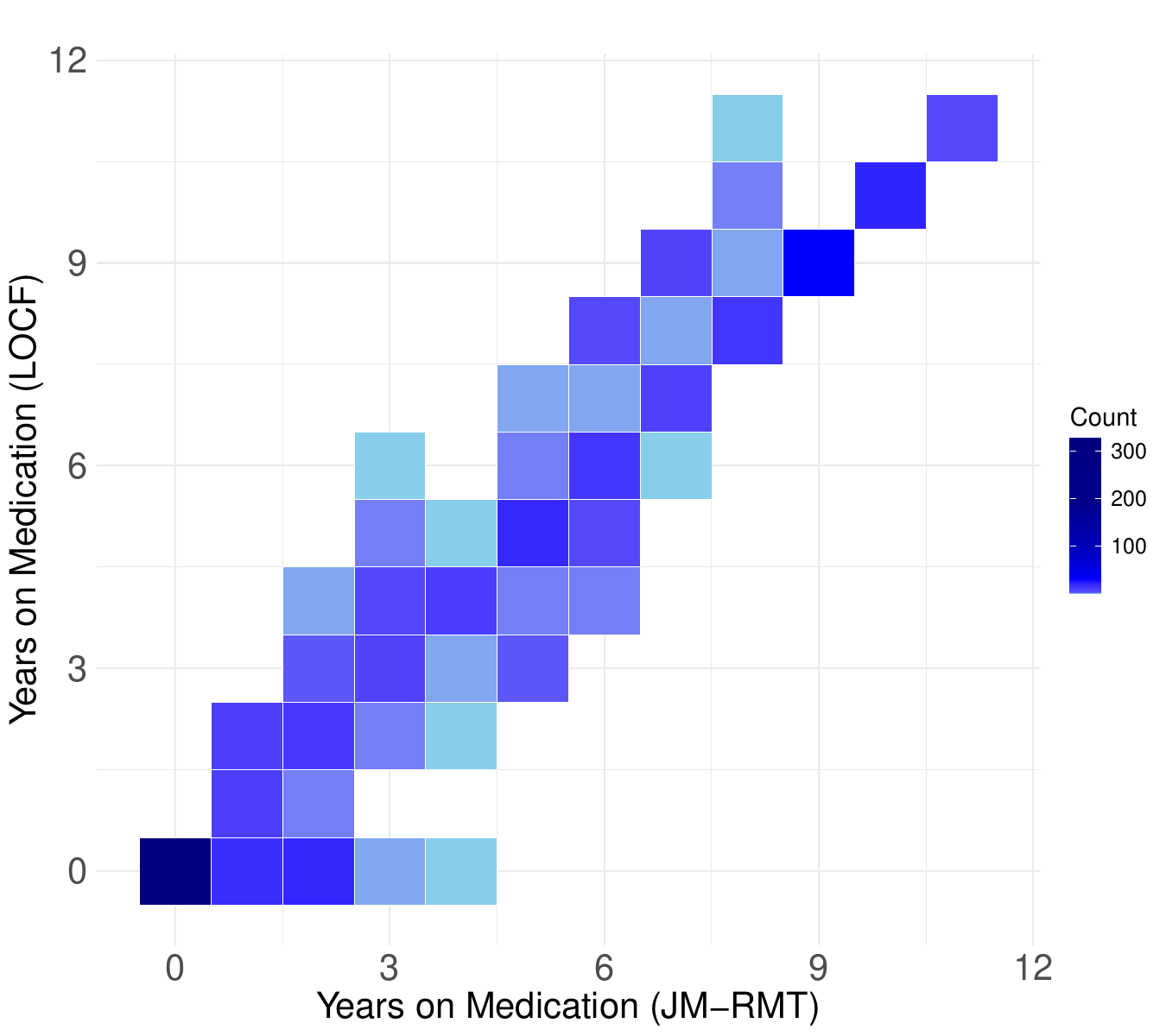}}
   \caption{Comparison of years on medication between JM-RMT and LOCF for men and women}
    \label{heatmap}
\end{figure}
Figure \ref{heatmap} shows the estimated years on medication under the JM-RMT and LOCF methods for men and women. For both sexes, the heatmaps show an evident concentration along the diagonal, indicating strong agreement between the methods for individuals with stable medication use. However, off-diagonal patterns highlight discrepancies where LOCF either overestimates or underestimates years on medication compared to JM-RMT. These differences are particularly evident for intermediate durations (3–6 years), where LOCF fails to account for potential medication switches during gaps between visits. Notably, the proportion of individuals with zero years on medication is lower in JM-RMT (men: 49.2\%, women: 54.0\%) compared to LOCF (men: 56.7\%, women: 60.9\%), indicating that JM-RMT identifies more instances of potential medication use. This suggests that LOCF’s simplistic assumptions may overclassify individuals as not being on medication. The patterns for women show greater spread, reflecting more variability in LOCF estimates compared to JM-RMT. These findings underscore the advantages of JM-RMT in capturing medication use dynamics and provides more realistic estimates by allowing transitions over time via the switch model in (\ref{stmodel}) and (\ref{stmodeldensity}).

While our results highlight these biases in observational data, it is important to note that in randomized controlled trials (RCTs), statins have been shown to provide a similar relative risk reduction (RRR) for cardiovascular disease events in both men and women, though women tend to have a lower absolute risk reduction (ARR) due to their lower baseline CVD risk.
%

\subsubsection{Predictive accuracy}
We expect the impact of medication on hazard to be more pronounced when using a precise measure of medication status, as opposed to a less accurate or noisy version. Measurement error in medication use is likely to attenuate the observed association between medication and the hazard \citep{carroll2006measurement}.
To assess predictive accuracy, we compare the hazard at individual death times under the JM-RMT model versus the LOCF method. This approach quantifies how frequently the hazard is greater under our model compared to LOCF. 

Based on a comparison, the posterior mean of the hazards at observed death times under our model exceeds that of the LOCF method in 88.6\% of men and 63.3\% of women. It provides strong evidence that our approach yields more accurate risk predictions.
To gain a probabilistic perspective, we extend this analysis to all posterior samples. By examining the distribution of 1000 posterior draws of hazards, we conduct an elementwise comparison. For each individual, we computed the proportion of the 1000 posterior draws in which our model estimated a higher hazard than LOCF.
This analysis reveals that for 88.6\% of men and 56.7\% of women, the proportion exceeds 0.5, further confirming that the JM-RMT model captures medication dynamics and their effects on outcomes more comprehensively than LOCF.

We also consider a comparison that excludes those who were always on or off medication with fully observed data, as the medication feature remains identical under both models in such cases. For this subset, our model estimates a higher posterior mean of hazard than the LOCF method in 81.2\% of men and 63.6\% of women. Based on the posterior draws comparison, 81.3\% of men and 72.7\% of women have proportions higher than 0.5. This highlights the model’s strength in accounting for medication variability.

\section{Discussion} \label{sec5}
Our proposed Bayesian approach for jointly modeling longitudinal cardiovascular disease (CVD) risk factor trajectories, medication use, and time-to-events offers a unified framework for leveraging observational data to better estimate the real-world effects of medications being used in non-randomized settings. By integrating these components, the methodology improves predictive accuracy and provides insights into how medication timing and duration influence CVD outcomes.
A key strength of the model is its ability to handle incomplete medication data by incorporating uncertainty in medication status and switch times. The explicit modeling of temporal dynamics allows the ability to accurately capture cumulative medication effects and their interactions with risk factor trajectories and CVD event hazards.

Our approach leverages shared parameters across components while incorporating time-varying features, making it well-suited for the complex nature of CVD progression. This flexibility enables a detailed exploration of the interplay between risk factors, medications, and events.

For future work, we could expand this model to include multiple risk factors and medications, enabling a more comprehensive understanding of their combined effects on CVD outcomes. 
Furthermore, integrating this approach into dynamic treatment strategies could optimize medication timing and combinations, ultimately improving long-term health outcomes.
\section*{Acknowledgements}
This work was supported by NIH/NHLBI 1R01HL158963 (MPI: Siddique/Daniels).

\section*{Supplementary Materials} 
The supplementary materials provide additional details on model specification and prior distributions. They include derivations of key equations and tables summarizing additional results. Further methodological justifications and robustness checks are presented to support the findings in the main text.


\begin{thebibliography}{}

\bibitem[\protect\citeauthoryear{Allen, Siddique, Wilkins, Shay, Lewis, Goff, Jacobs, Liu, and Lloyd-Jones}{Allen et~al.}{2014}]{allen2014blood}
Allen, N.~B., J.~Siddique, J.~T. Wilkins, C.~Shay, C.~E. Lewis, D.~C. Goff, D.~R. Jacobs, K.~Liu, and D.~Lloyd-Jones (2014).
\newblock Blood pressure trajectories in early adulthood and subclinical atherosclerosis in middle age.
\newblock {\em JAMA\/}~{\em 311\/}(5), 490--497.

\bibitem[\protect\citeauthoryear{Bild, Bluemke, Burke, Detrano, Diez~Roux, Folsom, Greenland, JacobsJr, Kronmal, Liu, et~al.}{Bild et~al.}{2002}]{bild2002multi}
Bild, D.~E., D.~A. Bluemke, G.~L. Burke, R.~Detrano, A.~V. Diez~Roux, A.~R. Folsom, P.~Greenland, D.~R. JacobsJr, R.~Kronmal, K.~Liu, et~al. (2002).
\newblock {Multi-Ethnic Study of Atherosclerosis: Objectives and design}.
\newblock {\em American Journal of Epidemiology\/}~{\em 156\/}(9), 871--881.

\bibitem[\protect\citeauthoryear{Carroll, Ruppert, Stefanski, and Crainiceanu}{Carroll et~al.}{2006}]{carroll2006measurement}
Carroll, R.~J., D.~Ruppert, L.~A. Stefanski, and C.~M. Crainiceanu (2006).
\newblock {\em Measurement Error in Nonlinear Models: A Modern Perspective}.
\newblock Chapman and Hall/CRC.

\bibitem[\protect\citeauthoryear{{Cholesterol Treatment Trialists' (CTT) Collaboration}}{{Cholesterol Treatment Trialists' (CTT) Collaboration}}{2010}]{trialists2010efficacy}
{Cholesterol Treatment Trialists' (CTT) Collaboration} (2010).
\newblock Efficacy and safety of more intensive lowering of {LDL} cholesterol: A meta-analysis of data from 170 000 participants in 26 randomised trials.
\newblock {\em The Lancet\/}~{\em 376\/}(9753), 1670--1681.

\bibitem[\protect\citeauthoryear{Crowther, Lambert, and Abrams}{Crowther et~al.}{2013}]{crowther2013adjusting}
Crowther, M.~J., P.~C. Lambert, and K.~R. Abrams (2013).
\newblock Adjusting for measurement error in baseline prognostic biomarkers included in a time-to-event analysis: A joint modelling approach.
\newblock {\em BMC Medical Research Methodology\/}~{\em 13}, 1--8.

\bibitem[\protect\citeauthoryear{{De Valpine}, Paciorek, Turek, Michaud, Anderson-Bergman, Obermeyer, Wehrhahn~Cortes, Rodr{\'i}guez, {Temple Lang}, Zhang, Paganin, Hug, and van Dam-Bates}{{De Valpine} et~al.}{2024}]{nimble-software:2024}
{De Valpine}, P., C.~Paciorek, D.~Turek, N.~Michaud, C.~Anderson-Bergman, F.~Obermeyer, C.~Wehrhahn~Cortes, A.~Rodr{\'i}guez, D.~{Temple Lang}, W.~Zhang, S.~Paganin, J.~Hug, and P.~van Dam-Bates (2024).
\newblock {NIMBLE: MCMC, Particle Filtering, and Programmable Hierarchical Modeling}.

\bibitem[\protect\citeauthoryear{Domanski, Wu, Tian, Hasan, Ma, Huang, Miao, Reis, Bae, Husain, et~al.}{Domanski et~al.}{2023}]{domanski2023association}
Domanski, M.~J., C.~O. Wu, X.~Tian, A.~A. Hasan, X.~Ma, Y.~Huang, R.~Miao, J.~P. Reis, S.~Bae, A.~Husain, et~al. (2023).
\newblock Association of incident cardiovascular disease with time course and cumulative exposure to multiple risk factors.
\newblock {\em Journal of the American College of Cardiology\/}~{\em 81\/}(12), 1151--1161.

\bibitem[\protect\citeauthoryear{Fitzmaurice, Laird, and Ware}{Fitzmaurice et~al.}{2012}]{fitzmaurice2012applied}
Fitzmaurice, G.~M., N.~M. Laird, and J.~H. Ware (2012).
\newblock {\em Applied longitudinal analysis}.
\newblock John Wiley \& Sons.

\bibitem[\protect\citeauthoryear{Hickey, Philipson, Jorgensen, and Kolamunnage-Dona}{Hickey et~al.}{2016}]{hickey2016joint}
Hickey, G.~L., P.~Philipson, A.~Jorgensen, and R.~Kolamunnage-Dona (2016).
\newblock Joint modelling of time-to-event and multivariate longitudinal outcomes: recent developments and issues.
\newblock {\em BMC Medical Research Methodology\/}~{\em 16}, 1--15.

\bibitem[\protect\citeauthoryear{Kostis, Cheng, Dobrzynski, Cabrera, and Kostis}{Kostis et~al.}{2012}]{kostis2012meta}
Kostis, W.~J., J.~Q. Cheng, J.~M. Dobrzynski, J.~Cabrera, and J.~B. Kostis (2012).
\newblock Meta-analysis of statin effects in women versus men.
\newblock {\em Journal of the American College of Cardiology\/}~{\em 59\/}(6), 572--582.

\bibitem[\protect\citeauthoryear{Laurie}{Laurie}{1997}]{laurie1997calculation}
Laurie, D. (1997).
\newblock {Calculation of Gauss-Kronrod quadrature rules}.
\newblock {\em Mathematics of Computation\/}~{\em 66\/}(219), 1133--1145.

\bibitem[\protect\citeauthoryear{Lewey, Shrank, Bowry, Kilabuk, Brennan, and Choudhry}{Lewey et~al.}{2013}]{lewey2013gender}
Lewey, J., W.~H. Shrank, A.~D. Bowry, E.~Kilabuk, T.~A. Brennan, and N.~K. Choudhry (2013).
\newblock Gender and racial disparities in adherence to statin therapy: a meta-analysis.
\newblock {\em American Heart Journal\/}~{\em 165\/}(5), 665--678.

\bibitem[\protect\citeauthoryear{Mihaylova, Emberson, Blackwell, Keech, Simes, Barnes, Voysey, Gray, Collins, Baigent, et~al.}{Mihaylova et~al.}{2012}]{mihaylova2012effects}
Mihaylova, B., J.~Emberson, L.~Blackwell, A.~Keech, J.~Simes, E.~Barnes, M.~Voysey, A.~Gray, R.~Collins, C.~Baigent, et~al. (2012).
\newblock The effects of lowering {LDL} cholesterol with statin therapy in people at low risk of vascular disease: meta-analysis of individual data from 27 randomised trials.
\newblock {\em Lancet (London, England)\/}~{\em 380\/}(9841), 581--590.

\bibitem[\protect\citeauthoryear{Pool, Ning, Wilkins, Lloyd-Jones, and Allen}{Pool et~al.}{2018}]{pool2018use}
Pool, L.~R., H.~Ning, J.~Wilkins, D.~M. Lloyd-Jones, and N.~B. Allen (2018).
\newblock Use of long-term cumulative blood pressure in cardiovascular risk prediction models.
\newblock {\em JAMA Cardiology\/}~{\em 3\/}(11), 1096--1100.

\bibitem[\protect\citeauthoryear{Rizopoulos and Ghosh}{Rizopoulos and Ghosh}{2011}]{rizopoulos2011bayesian}
Rizopoulos, D. and P.~Ghosh (2011).
\newblock A {Bayesian} semiparametric multivariate joint model for multiple longitudinal outcomes and a time-to-event.
\newblock {\em Statistics in medicine\/}~{\em 30\/}(12), 1366--1380.

\bibitem[\protect\citeauthoryear{Shao and Zhong}{Shao and Zhong}{2003}]{shao2003last}
Shao, J. and B.~Zhong (2003).
\newblock Last observation carry-forward and last observation analysis.
\newblock {\em Statistics in Medicine\/}~{\em 22\/}(15), 2429--2441.

\end{thebibliography}

\section*{Supplementary Materials} 
\subsection*{S.1 Gauss-Kronrod Quadrature} \label{S.1.}
We use Gauss-Kronrod quadrature with \( Q \) nodes to approximate the cumulative hazard
\[
\int_{a_{i1}}^{T_i} h_i(s) \, ds \approx \frac{T_i - a_{i1}}{2} \sum_{q=1}^{Q} w_q h_i\left(\frac{(T_i - a_{i1})(1 + s_q)}{2} + a_{i1}\right),
\]
where \( w_q \) and \( s_q \) respectively represent the standardized weights and locations ("abscissae") for quadrature node \( q \) (\( q = 1, \ldots, Q \)) \citep{laurie1997calculation}. 

To apply this method, medication status is assigned as follows: 
(i) for ages between 71 and 72, the medication status is recorded at age 71, and 
(ii) for ages between 72 and 73, the medication status is recorded at age 72. 
We choose to use \( Q = 15 \) quadrature nodes. Thus, for each individual with an event time \( T_i \), we evaluate the design matrices for the event submodel and longitudinal submodels at \( Q+1 \) time points. These points include the event time \( T_i \) and the quadrature points 
\[
\frac{(T_i - a_{i1})(1 + s_q)}{2} + a_{i1}, \quad q = 1, \ldots, Q.
\]
The log likelihood for the event submodel is then calculated by evaluating the hazard \( h_i(t) \) at these \( Q+1 \) time points and appropriately summing the contributions. This computation is performed during each iteration.

\subsection*{S.2 Prior Specification} \label{S.2.}
We use weakly informative priors. Independent normal priors with mean 0 and standard deviation 10 are placed on the regression coefficients for covariates across all submodels.
 To characterize the covariance structure of these random effects, we employ an inverse-Wishart prior for the covariance matrix \( \boldsymbol{\Sigma} \), specified as \( \boldsymbol{\Sigma} \sim \text{Inv-Wishart}(4, \boldsymbol{I}_{2}) \), where \( \boldsymbol{I}_{2} \) is the \( 2 \times 2 \) identity matrix. 
For the error scale parameter $\omega$, an inverse-gamma prior, $1/\omega \sim \text{Gamma}(0.01, 0.01)$, is used. 
A skewness parameter is set to 1.5 to account for potential asymmetry in the residuals.

\subsection*{S.3 Posterior Summaries of the Model Parameters} \label{S.3.}
\begin{table}[!t]
\centering
\caption{Posterior Summaries and Credible Intervals}
\begin{tabular}{lcccccccc}
\hline
 & \multicolumn{4}{c}{Men} & \multicolumn{4}{c}{Women} \\
 & Mean & SD & 2.5\% & 97.5\% & Mean & SD & 2.5\% & 97.5\% \\
\hline
&&&&&&&& \\ 
\emph{Longitudinal submodel} &&&&&&&& \\ 
Intercept & 159.32 & 2.1090 & 155.14 & 163.40 & 173.90 & 1.7770 & 170.50 & 177.43 \\
Edu (HS vs -HS)        & 2.4280 & 2.9270 & -3.3300 & 8.1680 & 4.9240 & 2.5590 & -0.0080 & 10.0390 \\
Edu (+HS vs -HS)         & 0.8210 & 2.3140 & -3.6450 & 5.4170 & 9.1460 & 2.1320 & 4.9840 & 13.3190 \\
Race (Black vs. non-Black) & -0.7191 & 2.0030 & -4.6192 & 3.2350 & -2.2550 & 1.8890 & -6.0000 & 1.4202 \\
Age        & -3.1412 & 0.5326 & -4.1872 & -2.1060 & -1.8430 & 0.4383 & -2.7090 & -1.0056 \\
Age:Edu (HS vs -HS)        & 0.8157 & 0.7038 & -0.5474 & 2.2090 & 1.0390 & 0.6012 & -0.1352 & 2.2339 \\
Age:Edu (+HS vs -HS) & 1.0913 & 0.5610 & 0.0049 & 2.2060 & 1.0850 & 0.5004 & 0.1155 & 2.0664 \\
Age:Race (Black vs. non-Black) & 1.4179 & 0.4506 & 0.5375 & 2.3060 & 0.2648 & 0.4323 & -0.5925 & 1.1082 \\
Age:Medication & -1.0119 & 0.3153 & -1.6333 & -0.3932 & -0.1630 & 0.3148 & -0.7691 & 0.4639 \\
$1/\omega$ & 0.0017 & 0.0000 & 0.0016 & 0.0018 & 0.0016 & 0.0000 & 0.0016 & 0.0017 \\
&&&&&&&& \\ 
\emph{Medication submodel} &&&&&&&& \\
$\alpha_{1}$       & -1.0330 & 0.4136 & -1.8340 & -0.2253 & -2.4300 & 0.4368 & -3.2850 & -1.5683 \\
$\alpha_{2}$       & 5.4180 & 0.7585 & 3.9471 & 6.9450 & 7.9650 & 0.8096 & 6.3950 & 9.5374 \\
$\alpha_{3}$       & 0.1414 & 0.0251 & 0.0922 & 0.1906 & 0.0623 & 0.0217 & 0.0201 & 0.1049 \\
$\alpha_{4}$       & -0.1482 & 0.0437 & -0.2345 & -0.0630 & -0.0977 & 0.0393 & -0.1756 & -0.0210 \\
$\alpha_{5}$       & -0.1850 & 0.5725 & -1.3310 & 0.9205 & -1.4630 & 0.5529 & -2.5480 & -0.3843 \\
$\alpha_{6}$       & 1.6337 & 1.0480 & -0.4281 & 3.6880 & 2.0390 & 0.9615 & 0.1580 & 3.9310 \\
$\alpha_{7}$       & -0.0053 & 0.0023 & -0.0099 & -0.0008 & 0.0043 & 0.0022 & -0.0002 & 0.0087 \\
$\alpha_{8}$       & -0.0124 & 0.0043 & -0.0210 & -0.0041 & -0.0262 & 0.0042 & -0.0343 & -0.0180 \\
&&&&&&&& \\ 
\emph{Survival submodel} &&&&&&&& \\
Edu (HS vs -HS)  & 0.2610 & 0.4427 & -0.6131 & 1.1260 & 0.3807 & 0.5613 & -0.7502 & 1.4590 \\
Edu (+HS vs -HS)  & 0.1567 & 0.3387 & -0.4803 & 0.8545 & 0.2234 & 0.4610 & -0.6640 & 1.1492 \\
Race (Black vs. non-Black) & 0.3836 & 0.3479 & -0.3245 & 1.0390 & 0.1151 & 0.4768 & -0.8581 & 1.0158 \\
\hline
\end{tabular}
\label{JMRMTresults}
\end{table}
Table \ref{JMRMTresults} provides the posterior summaries of the model parameters. 
In the longitudinal submodel, parameters associated with age and its interactions with education and race reveal significant associations with risk factor trajectories. For instance, the interaction term Age:Race (Black vs. non-Black) indicates that Black individuals may experience distinct age-related changes in risk factors compared to non-Black individuals. The negative coefficient for Age:Medication suggests that the decline in risk factor levels with increasing age is more pronounced among individuals using medication.
%

Recall the medication model (\ref{medmodel})
\begin{eqnarray}
   && \text{logit}\left[\Pr \big(m_{i}(a_{ij}) = 1 \mid m_{i}(a_{i,j-1}) = m, g^{\mu}(\mu_{i}(\ell); \ell \leq a_{i,j-1})\big) \right] \notag \\
   &&= \gamma_{0}^{(m)}(a_{ij}) + \gamma_{1}^{(m)}(a_{ij} - a_{i,j-1}) + \gamma_{2}^{(m)}g^{\mu}(\mu_{i}(\ell); \ell \leq a_{i,j-1}), 
\end{eqnarray}
where \( \ell \) represents ages less than or equal to \( a_{i,j-1} \), \( \mu_{i}(\ell) \) denotes the true risk factor trajectory, and \( g^{\mu}(\mu_{i}(\ell); \ell \leq a_{i,j-1}) \) refers to the risk factor feature up to age \( a_{i,j-1} \). Here, \( m \) represents the previous medication status. 
The baseline intercept, \( \gamma_0^{(m)} \), was reparameterized as a linear function of age (\( a_{ij} \)) to enhance interpretability. Specifically, \( \gamma_0^{(0)} = \alpha_1 + \alpha_3 a_{ij} \) represents the effect of age when the previous medication status is \( m = 0 \), while \( \gamma_0^{(1)} = (\alpha_1 + \alpha_2) + (\alpha_3 + \alpha_4) a_{ij} \) captures the effect of age when the previous medication status is \( m = 1 \).
The parameter \( \gamma_1^{(m)} \), which accounts for the influence of the time gap between visits (\( |a_{ij} - a_{i,j-1}| \)), was modeled using an exponential decay function to reflect the gradual reduction of its influence over time. Specifically, \( \gamma_1^{(0)} = \alpha_5 \exp(-|a_{ij} - a_{i,j-1}|) \) when \( m = 0 \), and \( \gamma_1^{(1)} = (\alpha_5 + \alpha_6) \exp(-|a_{ij} - a_{i,j-1}|) \) when \( m = 1 \).
Finally, \( \gamma_2^{(m)} \), which represents the effect of the risk factor feature in the medication model, was parameterized as \( \gamma_2^{(0)} = \alpha_7 \) for \( m = 0 \), and \( \gamma_2^{(1)} = \alpha_7 + \alpha_8 \) for \( m = 1 \). Then, we have the full representation of the model as
\begin{eqnarray}
&& \text{logit}\left[\Pr \big(m_{i}(a_{ij}) = m \mid m_{i}(a_{i,j-1}) =  m, g^{\mu}(\mu_{i}(\ell); \ell \leq a_{i,j-1})\big) \right] = \notag \\
~&& \big(\alpha_1 + \alpha_2 m\big) + \big(\alpha_3 + \alpha_4 m\big) a_{ij} + \big(\alpha_5 + \alpha_6 m\big) \exp(-|a_{ij} - a_{i,j-1}|) + \big(\alpha_7 + \alpha_8 m\big) g^{\mu}(\mu_{i}(\ell); \ell \leq a_{i,j-1}). \notag 
\end{eqnarray}
%
The posterior estimates indicate gender differences in medication use dynamics. Baseline age (\( \alpha_1 \)) negatively affects medication use, with the effect stronger in women (\(-2.43\)) than men (\(-1.03\)). Prior medication (\( \alpha_2 \)) strongly predicts continued use, more so for women (\(7.97\)) than men (\(5.42\)). Age (\( \alpha_3 \)) has a small positive effect, more pronounced in men, while its interaction with prior medication (\( \alpha_4 \)) is negative for both, stronger in men. Risk factor effects (\( \alpha_7 \)) are slightly negative for men and positive for women, while its interaction with prior medication (\( \alpha_8 \)) is negative, stronger in women.
In the survival submodel, the effects of education and race on hazard rates appear less pronounced, as the credible intervals for most estimates span zero. 
\end{document}